\documentclass[oribibl]{llncs}

\usepackage{amsmath}
\usepackage{amsfonts}
\usepackage{proof}
\usepackage{stmaryrd}
\usepackage{enumerate}
\usepackage{Tabbing}
\usepackage{hyperref}
\usepackage{syntax}
\usepackage{fancyvrb}
\usepackage{pgfplots}

\RequirePackage{array}
\RequirePackage{vk}
\usepackage{cleveref} 
\crefname{section}{Sect.}{Sect(s).}
\crefname{definition}{Def.}{Def(s).}
\crefname{figure}{Fig.}{Fig(s).}
\crefname{proposition}{Prop.}{Prop(s).}
\crefname{lemma}{Lem.}{Lem.(s)}
\crefname{theorem}{Thm.}{Thm(s).}
\crefname{corollary}{Cor.}{Cor(s).}
\crefname{examplethm}{Ex.}{Ex(s).}
\crefname{appendix}{Appendix}{Appendices}
\crefname{statement}{Stmt.}{Stmt(s).}

\newtheorem{examplethm}[theorem]{Example}

\font\tt=rm-lmtl10
-lmtlo10
\font\btt=rm-lmtk10
-lmtko10
\makeatletter
\def\verbatim@font{\tt}
\makeatother

\newcommand{\lang}{\text{TCML}\xspace}
\newcommand{\defn}[1]{\textsf{#1}\xspace}
\newcommand{\Alice}{\defn{Alice}}
\newcommand{\Bob}{\defn{Bob}}
\newcommand{\Carol}{\defn{Carol}}
\newcommand{\David}{\defn{David}}

\newcommand{\src}[1]{{\tt #1}\xspace}
\newcommand{\movie}{\src{movie}}
\newcommand{\dinner}{\src{dinner}}
\newcommand{\dancing}{\src{dancing}}

\newcommand{\keyw}[1]{{\btt #1}\xspace}
\newcommand{\sync}{\keyw{sync}}

\newcommand{\ident}[1]{\textit{#1}}

\usepackage{tikz}
\usetikzlibrary{decorations.pathmorphing}     
\usetikzlibrary{matrix}                       
\usetikzlibrary{arrows, decorations.markings} 
\usetikzlibrary{calc}                         
\usetikzlibrary{automata}
\usetikzlibrary{calc, trees, positioning, arrows, shapes, shapes.multipart, 
shadows, matrix, decorations.pathreplacing, decorations.pathmorphing}

 \DefineVerbatimEnvironment{code}{Verbatim}{fontsize=\small}
\DefineVerbatimEnvironment{example}{Verbatim}{fontsize=\small}
\newcommand{\ignore}[1]{}

\newenvironment{nstabbing}
  {\setlength{\topsep}{-\parskip}%
   \setlength{\partopsep}{0pt}%
   \tabbing}
  {\endtabbing}

\usepackage{./rules}
\usepackage{./TCML}

\begin{document}

\frontmatter          

\pagestyle{headings}  

\mainmatter            
\title{Towards Efficient Abstractions for Concurrent Consensus\thanks{%
    Student project paper (primarily the work of the first author).
}}
\titlerunning{\toctitle}  
%
\author{%
  Carlo Spaccasassi\thanks{Supported by MSR (MRL 2011-039)}
  \and
  Vasileios Koutavas\thanks{Supported by SFI project SFI 06 IN.1 1898.}
}

\authorrunning{C.\ Spaccasassi \and V.\ Koutavas} 
%
%
\institute{Trinity College Dublin, Ireland\\
\email{\{spaccasc,Vasileios.Koutavas\}@scss.tcd.ie}}

\maketitle              

\begin{abstract}
Consensus is an often occurring problem in concurrent and distributed
programming. We present a programming language with simple semantics and
build-in support for consensus in the form of communicating transactions. We
motivate the need for such a construct with a characteristic example of
generalized consensus which can be naturally encoded in our language.
We then focus on the challenges in achieving an implementation that can
efficiently run such programs. We setup an architecture to evaluate different
implementation alternatives and use it to experimentally evaluate runtime
heuristics. This is the basis for a research project on realistic programming
language support for consensus.

\end{abstract}
\section{Introduction}
    
Achieving consensus between concurrent processes is a ubiquitous problem in
multicore and distributed programming \cite{distribComp,herlihybook}. Among the
classic instances of consensus is leader election and synchronous multi-process
communication. Programming language support for consensus, however, has been
limited. For example, CML's first-class communication primitives provide a
programming language abstraction to implement two-party consensus. However, they
cannot be used to abstractly implement consensus between three or more processes
\cite[Thm.\ 6.1]{reppy}---this needs to be implemented in a case-by-case basis.

Let us consider a hypothetical scenario of generalized consensus, which we will
call the \emph{Saturday Night Out} (SNO) problem. In this scenario a number of
friends are seeking partners for various activities on Saturday night.
Each has a list of desired activities to attend in a certain order, and will only
agree for a night out if there is a partner for each activity. \Alice, for
example, is looking for company to go out for dinner and then a movie (not
necessarily with the same person). To find partners for these events in this
order she may attempt to synchronize on the ``handshake'' channels \dinner and \movie:
\begin{alltt}
  \Alice \defeq \sync dinner; \sync movie
\end{alltt}
Here \sync is a two-party synchronization operator, similar to CSP
synchronization.
\Bob, on the other hand, wants to go for dinner and then for dancing:
\begin{alltt}
  \Bob \defeq \sync dinner; \sync dancing
\end{alltt}
\Alice and \Bob can agree on dinner but they need partners for a movie and
dancing, respectively, to commit to the night out. Their agreement is
\emph{tentative}.

Let \Carol be another friend in this group who is only interested in dancing:
\begin{alltt}
  \Carol \defeq \sync dancing
\end{alltt}
Once \Bob and \Carol agree on dancing they are both happy to commit to going
out. However, \Alice has no movie partner and she can still cancel her agreement
with \Bob. If this happens, \Bob and \Carol need to be notified to cancel their
agreement and everyone starts over their search of partners. An implementation
of the SNO scenario between concurrent processes would need to have a
specialized way of reversing the effect of this synchronization. Suppose \David is also
a participant in this set of friends.
\begin{alltt}
  \David \defeq \sync dancing; \sync movie
\end{alltt}
After the partial agreement between \Alice, \Bob, and \Carol is canceled,
\David together with the first two can synchronize on \dinner,
\dancing, and \movie and agree to go out (leaving \Carol at home).

Notice that when \Alice raised an objection to the agreement that was forming
between her, \Bob, and \Carol, all three participants were forced to restart.
If, however, \Carol was taken out of the agreement (even after she and \Bob were
happy to commit their plans), \David would have been able to take \Carol's place
and the work of \Alice and \Bob until the point when \Carol joined in
would not need to be repeated.

Programming SNO between an arbitrary number of processes (which can form multiple
agreement groups) in CML is complicated. Especially if we consider that the
participants are allowed to perform arbitrary computations between
synchronizations affecting control flow, and can communicate with other parties
not directly involved in the SNO. For example, \Bob may want to go dancing only
if he can agree with the babysitter to stay late:
\begin{alltt}
  \Bob \defeq \sync dinner; \keyw{if} babysitter() \keyw{then} \sync dancing
\end{alltt}
In this case \Bob's computation has side-effects outside of the SNO group of
processes. To implement this would require code for dealing with the SNO
protocol to be written in the \defn{Babysitter} (or any other) process, breaking
any potential modular implementation.

This paper shows that \emph{communicating transactions}, a recently proposed
mechanism for automatic error recovery in CCS processes \cite{commTransactions},
is a useful mechanism for modularly implementing the SNO and other generalized
consensus scenarios. Previous work on communicating transactions focused on
behavioral theory with respect to \emph{safety} and \emph{liveness}
\cite{commTransactions,livenessComm}. However, the effectiveness of this
construct in a pragmatic programming language has yet to be proven. One of the
main milestones to achieve on this direction is the invention of efficient
runtime implementations of communicating transactions. Here we describe the
challenges and our first results in a recently started project to investigate
this research direction.

In particular, we equip a simple concurrent functional language with
communicating transactions and use it to discuss the challenges in making
an efficient implementation of such languages (\cref{section:language}). We also
use this language to give a modular implementation of consensus scenarios such
as the SNO example. The simple operational semantics of
this language allows for the communication of SNO processes with arbitrary other
processes (such as the \defn{Babysitter} process) without the need to add code
for the SNO protocol in those processes. Moreover, the more efficient, partially
aborting strategy discussed above is captured in this semantics.

Our semantics of this language is non-deterministic, allowing different runtime
scheduling strategies of processes, some more efficient than others. To study
their relative efficiency we have developed a skeleton implementation of the
language which allows us to plug in and evaluate such runtime strategies
(\cref{section:implementation}). We describe several such strategies
(\cref{section:strategies}) and report the results of our evaluations
(\cref{section:experiments}). Finally, we summarize related work in this area
and the future directions of this project (\cref{section:related-work}).


\section{The \lang Language}                        \label{section:language}
    \label{syntax}

\begin{myfigure*}{\lang syntax.}{tab:syntaxTable} 

  $$\begin{array}{@{}r@{~}c@{~}l}
    \ \\
    A & ::= & 
              \Unitb
      \bnf    \Boolb
      \bnf    \Intb
      \bnf    A \starb A
      \bnf    A \arrowb A
      \bnf    A \Chanb
    \\[1ex]
    v 
    & ::= 
    &
      x
      \bnf ()
      \bnf \trueb
      \bnf \falseb
      \bnf   n 
      \bnf
                  (v, v)
            \bnf  \funco{f}{x}{e}
      \bnf  c   
    \\[1ex]
    e 
    & ::= 
    &        v 
        \bnf  (e, e)  
        \bnf  e\; e 
        \bnf  op\; e 
      
     \bnf  \letb x = e \inb e  \bnf  \ifb e \thenb e \elseb e
    \\
    & \bnf
    &      \sendo{e}{e} 
      \bnf  \recvo{e} 
      \bnf  \newchanb_A 
      \bnf  \spawnb e
    \\
    &  \bnf 
    &  \atomico{k}{e}{e} \bnf \commitb k 
    \\[1ex]
     P  
    &  ::= 
    &        e  
        \bnf  P \parallel P  
        \bnf  \nub c. P  
        \bnf  \lt P \rhd_k P \rt 
        \bnf  \cob k 
    \\[1ex]
     op  & ::= & 
    
            \fstb
      \bnf  \sndb 
      \bnf  \addb
      \bnf  \subb
      \bnf  \mulb
      \bnf  \leqb
   \\[1ex]
     E
    & ::=   
    & 
          [] 
      \bnf (E, e) 
      \bnf (v, E) 
      \bnf E\; e 
      \bnf v\; E
      \bnf op\; E 
      \bnf \letb x = E \inb e
   \\&\bnf& \ifb E \thenb e_1 \elseb e_2
      \bnf \sendo{E}{e}
      \bnf \sendo{v}{E}
      \bnf \recvo{E} 
      \bnf \spawnb E
   \\[1ex]
   \multicolumn{3}{@{}l@{}}{\text{
    where $ n \in \mathbb{N} $, $ x \in \mathit{Var} $, $ c \in \mathit{Chan} $, $k \in \mathcal{K} $  
   }}\\[0.7ex]
   \end{array}$$

\end{myfigure*} 

We study \lang, a language combining a simply-typed $\lambda$-calculus with
$\pi$-calculus and communicating transactions. For this language we use the
abstract syntax shown in \cref{tab:syntaxTable} and the usual abbreviations from
the $\lambda$- and $\pi$-calculus.
Values in \lang are either constants of base type ($\Unitb$, $\Boolb$, and
$\Intb$), pairs of values (of type $A \starb A$), recursive functions ($A
\arrowb A$), and channels carrying values of type $A$ ($A \Chanb$). A simple type
system (with appropriate progress and preservation theorems) can be found in an
accompanying technical report \cite{tcml-report} and is omitted here.

Source \lang programs are expressions in the functional core of the language,
ranged over by $e$, whereas running programs are processes derived from the
syntax of $P$. 
Besides standard lambda calculus expressions, the functional core
contains the constructs $\sendo{c}{e}$ and $\recvo{c}$ to synchronously send and
receive a value on channel $c$, respectively, and $\newchanb_A$ to create a new
channel of type $\Chanb A$. The constructs $\spawnb$ and $\atomicb$, when
executed, respectively spawn a new process and transaction; $\commitb k$
commits transaction $k$---we will shortly describe these constructs in detail.

A simple running process can be just an expression $e$.
It can also be constructed by the parallel composition of $P$ and $Q$ ($P
\parallel Q$). We treat free channels as in the $\pi$-calculus, considering them
to be \emph{global}. Thus if a channel $c$ is free in both 
$P$ and $Q$, it can be used for communication between these processes.
The construct $\nub c. P$ encodes $\pi$-calculus restriction of the scope of $c$
to process $P$. We use the Barendregt convention for bound variables and
channels and identify terms up to alpha conversion. Moreover, we write $\fc(P)$
for the free channels in process $P$.

We write $\lt P_1 \rhd_k P_2 \rt$ for the process encoding a communicating
transaction. This can be thought of as the process $P_1$, the \emph{default} of the
transaction, which runs until the transaction \emph{commits}. If, however, the
transaction \emph{aborts} then $P_1$ is discarded and the entire transaction is
replaced by its \emph{alternative} process $P_2$. Intuitively, $P_2$ is the
continuation of the transaction in the case of an abort.
As we will explain, commits are asynchronous, requiring the addition of process
$\cob k$ to the language. The name $k$ of the transaction is bound in $P_1$.
Thus only the default of the transaction can potentially spawn a $\cob k$. The
meta-function $\ftn(P)$ gives us the free transaction names in $P$.

\begin{myfigure*}{Sequential reductions}{tab:sequentialReductionRules} 
  $$\begin{array}{llcll}
     \EIfT &
        \ifb \trueb \thenb e_1 \elseb e_2 &\hookrightarrow &e_1
    \\
     \EIfF &
      \ifb \falseb \thenb e_1 \elseb e_2 &\hookrightarrow &e_2
    \\
     \ELet &
       \letb x = v \inb e &\hookrightarrow &e[v/x] 
    \\
     \EOp &
      \opb v &\hookrightarrow &\delta(\opb, v)
    \\
     \EApp &
      \appo{\funco{f}{x}{e}}{v_2} &\hookrightarrow &e[\funco{f}{x}{e}/f][v_2/x]
\\[1em]
      \CStep &
      E[e] &\arrow& E[e']
      &\text{if~~}
       e \hookrightarrow e'
    \\
      \CSpawn &
      E[\spawnb v] &\arrow& v\; () \parallel E[()]
   	\\
      \CNewChan &
       E[\newchanb_A ] &\arrow& \nu c. E[c]
       &\text{if~~}
        c \not \in \fc(E[()])
   \\
    \TrAtomic &
      E [\atomico{k}{e_1}{e_2}] &\arrow& \transo{k}{E[e_1]}{E[e_2]} 
    \\
     \TrCommit &
    E[\commitb k] &\arrow& \cob k \parallel E[()]
   \end{array}$$
\end{myfigure*} 

Processes with no free variables can reduce using transitions of the form $P
\arrow Q$. These transitions for the functional part of the language are shown
in \cref{tab:sequentialReductionRules} and are defined in terms of reductions $e
\hookrightarrow e'$ (where $e$ is a \emph{redex}) and eager, left-to-right
evaluation contexts $E$ whose grammar is given in \cref{tab:syntaxTable}. Due to
a unique decomposition lemma, an expression $e$ can be decomposed to an
evaluation context and a redex expression in only one way.
Here we use $e[u/x]$ for the standard capture-avoiding substitution, and
$\delta(\mathit{op}, v)$ for a meta-function returning the result of the operator
$\mathit{op}$ on $v$, when this is defined.

Rule $\CStep$ lifts functional reductions to process reductions. The rest of the
reduction rules of \cref{tab:sequentialReductionRules} deal with the
concurrent and transactional side-effects of expressions.
Rule $\CSpawn$ reduces a $\spawnb v$ expression at evaluation position to the
unit value, creating a new process running the application $v\;()$. The type
system of the language guarantees that value $v$ here is a thunk. With this
rule we can derive the reductions:
$$\abox{
  \spawnb (\funco{f}{}{\sendo{c}{1}});
  \recvo{c}
  &\arrow
  (\funco{f}{}{\sendo{c}{1}})\; () \parallel \recvo{c}
  \\&\arrow
  \sendo{c}{1} \parallel \recvo{c}
}$$
The resulting processes of these reductions can then communicate on channel $c$.
As we previously mentioned, the free channel $c$ can also be used to communicate
with any other parallel process. Rule $\CNewChan$ gives processes the ability to
create new, locally scoped channels. Thus, the following expression will result
in an input and an output process that can \emph{only} communicate with each other:
$$\nbox{
  \leto{x}{ 
    {\newchanb}_{\Intb}
  }{(
    \spawno{ (\funco{f}{}{\sendo{x}{1}}) };
    \recvo{x}
  )}
  \\\arrow
    \nu c.\left(
    \spawno{ (\funco{f}{}{\sendo{c}{1}}) };
    \recvo{c} \right)
  \\\arrow^*
    \nu c.\left(
     \sendo{c}{1} 
     \parallel
    \recvo{c} \right)
}$$

Rule $\TrAtomic$ starts a new transaction in the current (expression-only) process, engulfing the
entire process in it, and storing the abort continuation in the alternative of
the transaction. Rule $\TrCommit$ spawns an asynchronous commit.
Transactions can be arbitrarily nested, thus we can write:
$$\nbox{
  \atomico{k}{\spawnb (\lambda(). \recvb c; \commitb k)}{()};~
    \\ \atomico{l}{\recvb d; \commitb l}{()}
}$$
which reduces to
$$\nbox{
  \transo{k}{
    (\recvb c; \commitb k)
    \parallel
    \transo{l}{\recvb d; \commitb l}{()}
    \\
  }{ 
    (); \atomico{l}{\recvb d; \commitb l}{()}
  }
}$$
This process will commit the $k$-transaction after an input on channel $c$ and
the inner $l$-transaction after an input on $d$. As we will see, if the $k$
transaction aborts then the inner $l$-transaction will be discarded (even if
it has performed the input on $d$) and the resulting process (the alternative of
$k$) will restart $l$:
$$
    (); \atomico{l}{\recvb d; \commitb l}{()}
$$
The effect of this abort will be the rollback of the communication on $d$
reverting the program to a consistent state.

\begin{myfigure*}{Concurrent and Transactional reductions (omitting symmetric rules).}{tab:concurrencyReductionRules} 
  $$\begin{array}{@{}l@{\qquad}l@{}}
    \linfer[ \CSync ]
        {}
        { E_1[\recvo{c}] \parallel E_2[\sendo{c}{v}] \arrow E_1[v]\parallel E_2[()]}
    &
    \linfer [ \CEq ]
        {P \equiv P' \arrow Q' \equiv Q}
        {P \arrow Q}
    \\[1.5em]
    \linfer [\CPar]
      { P_1 \arrow P_1'}
      { P_1 \parallel P_2 \arrow P_1' \parallel P_2}
    &
    \linfer [\CChan]
      {P \arrow P'}
      {\nu c. P \arrow \nu c.P'}
    \\[1.5em]
    \linfer  [\TrEmb ]
       {}
       { P_1 \parallel \transo{k}{P_2}{P_3} \arrow 
        \transo{k}{P_1 \parallel P_2}{P_1 \parallel  P_3} }
    &
    \linfer [ \TrStep ]
      { P \arrow P' }
      { \transo{k}{P}{P_2} \arrow \transo{k}{P'}{P_2} }
    \\[1.5em]
    \linfer   [ \TrCo ]
      { P_1 \equiv \cob k \parallel P_1' } 
      { \lt P_1 \rhd_k P_2 \rt \arrow P_1' \slash k }
    &
    \linfer   [\TrAbort ]
      {}
      {\lt P_1 \rhd_k P_2 \rt     \arrow     P_2 }
  \end{array}$$
\end{myfigure*} 

Process and transactional reductions are handled by the rules of
\cref{tab:concurrencyReductionRules}. The first four rules ($\CSync$, $\CEq$,
$\CPar$, and $\CChan$) are direct adaptations of the reduction rules of the
$\pi$-calculus, which allow parallel processes to communicate, and propagate
reductions over parallel and restriction. These rules use an omitted structural
equivalence $(\equiv)$ to identify terms up to the reordering of parallel
processes and the extrusion of the scope of restricted channels, in the spirit
of the $\pi$-calculus semantics. Rule $\TrStep$ propagates reductions of default
processes over their respective transactions. The remaining rules are
taken from TransCCS \cite{commTransactions}.

Rule $\TrEmb$ encodes the $\emph{embedding}$ of a process $P_1$ in a parallel
transaction $\lt P_1 \rhd_k P_2 \rt$. This enables the communication of $P_1$
with $P_2$, the default of $k$. It also keeps the current continuation of $P_1$ in
the alternative of $k$ in case the $k$-transaction aborts.
To illustrate the mechanics of the embed rule, let us consider
the above nested transaction running in parallel with the process $P=
\sendo{d}{()}; \sendo{c}{()}$:
\label{example:nested}
$$\nbox{
  \transo{k}{
    (\recvb c; \commitb k)
    \parallel
    \transo{l}{\recvb d; \commitb l}{()}
    \\
  }{ 
    (); \atomico{l}{\recvb d; \commitb l}{()}
  } \qquad\quad  \parallel \ \ P
}$$
After two embedding transitions we will have
$$\nbox{
  \transo{k}{
    (\recvb c; \commitb k)
    ~~\parallel~~
    \transo{l}{P \parallel \recvb d; \commitb l~}{~P \parallel ()}
  ~~}{~~ 
    P \parallel \dots
  }
}$$
Now $P$ can communicate on $d$ with the inner transaction:
$$\nbox{
  \transo{k}{
    (\recvb c; \commitb k)
    ~~\parallel~~
    \transo{l}{ \sendo{c}{()} \parallel \commitb l~}{~P \parallel ()}
  ~~}{~~ 
    P \parallel \dots
  }
}$$
Next, there are (at least) two options: either $\commitb l$ spawns a $\cob l$
process which causes the commit of the $l$-transaction, or the input on $d$ is
embedded in the $l$-transaction. Let us assume that the latter occurs:
$$\nbox{
  \transo{k}{~\nbox{
    \transo{l}{
      (\recvb c; \commitb k) ~\parallel~ \sendo{c}{()} ~\parallel~ \commitb l
    \\}{
      (\recvb c; \commitb k) ~\parallel~ P ~\parallel~ ()}
  }\\}{
    P \parallel \dots
  }
  \\\arrow^*
  \transo{k}{
    \transo{l}{
      \cob k \parallel \cob l
    }{
      \dots
    }
  }{
    \dots
  }
}$$
The transactions are now ready to commit from the inner-most to the outer-most
using rule $\TrCommit$. Inner-to-outer commits are necessary to guarantee that all transactions
that have communicated have reached an agreement to commit. 

This also has the important consequence of making the following three processes
behaviorally indistinguishable:
$$
  \transo{k}{P_1}{P_2} \parallel \transo{l}{Q_1}{Q_2}
$$
$$
  \transo{k}{
    P_1 \parallel 
    \transo{l}{
      Q_1
    }{
      Q_2
    }
  }{
    P_2
    \parallel
    \transo{l}{
      Q_1
    }{
      Q_2
    }
  }
$$
$$
  \transo{l}{
    \transo{k}{
      P_1
    }{
      P_2
    }
    \parallel Q_1
  }{
    \transo{k}{
      P_1
    }{
      P_2
    }
    \parallel
    Q_2
  }
$$
Therefore, an implementation of \lang, when dealing with the first of the three
processes can pick any of the alternative, non-deterministic mutual embeddings
of the $k$ and $l$ transactions without affecting the observable outcomes of the
program. In fact, when one of the transactions has no
possibility of committing or when the two transactions never communicate, an
implementation can decide \emph{never} to embed the two transactions in
each-other.
This is crucial in creating implementations that will only embed processes (and
other transactions) only when necessary for communication, and pick the most
\emph{efficient} of the available embeddings. The development of implementations
with efficient embedding strategies is one of the main challenges of our project
for scaling communicating transactions to pragmatic programming languages.

Similarly, aborts are entirely non-deterministic ($\TrAbort$) and are left to
the discretion of the underlying implementation. Thus in the above example any
transaction can abort at any stage, discarding part of the computation.
In such examples there is usually a multitude of transactions that can be
aborted, and in cases where a ``forward'' reduction is not possible (due to
deadlock) aborts are necessary. Making the \lang programmer in charge of aborts
(as we do with commits) is not desirable since the purpose of communicating
transactions is to lift the burden of manual error prediction and handling.
Minimizing aborts, and automatically picking the aborts that will undo the fewer
computation steps while still rewinding the program back enough to reach a
successful outcome is another major challenge in our project.

The SNO scenario can be simply implemented in \lang using \emph{restarting
transactions}. A restarting transaction uses recursion to re-initiate an
identical transaction in the case of an abort:
$$
  \restarto{k}{e} ~~\defeq~~
  \funco{r}{{}}{\atomico{k}{e}{\appo{r}()}}
$$
A transactional implementation of the SNO participants we discussed in the
introduction simply wraps their code in restating transactions:

\begin{center}
\begin{nstabbing}
  $
  \leto{\ident{alice} \quad $\=$}{\restarto{k}{
  \appo{\ident{sync}}{\ident{dinner}};\ \appo{\ident{sync}}{\ident{movie}};\ \commitb k}}{$\\$
    \leto{\ident{bob}$\>$}{\restarto{k}{
      \appo{\ident{sync}}{\ident{dinner}};\ 
  \appo{\ident{sync}}{\ident{dancing}} ;\ \commitb k}}{$\\$
    \leto{\ident{carol}$\>$}{\restarto{k}{
    \appo{\ident{sync}}{\ident{dancing}};\ \commitb k}}{$\\$
      \leto{\ident{david}$\>$}{\restarto{k}{
      \appo{\ident{sync}}{\ident{dancing}};\ \appo{\ident{sync}}{\ident{movie}}; \ \commitb k}}{$\\$
  \spawno{\ident{alice}};\ \spawno{\ident{bob}};\ \spawno{\ident{carol}};\ \spawno{\ident{david}} }} }}
  $
\end{nstabbing} 
\end{center}

Here $\ident{dinner}$, $\ident{dancing}$, and $\ident{movie}$ are
implementations of CSP synchronization channels and $\ident{sync}$ a function to
synchronize on these channels. Compared to a potential ad-hoc implementation of
SNO in CML the simplicity of the above code is evident (the version of \Bob
communicating with the \defn{Babysitter} is just as simple). However, as we discuss
in \cref{section:experiments}, this simplicity comes with a severe performance
penalty, at least for straightforward implementations of \lang. In essence, the
above code asks from the underlying transactional implementation to solve an
NP-complete satisfiability problem. Leveraging existing useful heuristics for
such problems is something we intend to pursue in future work.

In the following sections we describe an implementation where these
transactional scheduling decisions can be plugged in, and a number of heuristic
transactional schedulers we have developed and evaluated. Our work shows that
although more advanced heuristics bring measurable performance benefits, the
exponential number of runtime choices require the development of innovative
compilation and execution techniques to make communicating transactions a
realistic solution for programmers.


\section{An Extensible Implementation Architecture}     \label{section:implementation}
  We have developed an interpreter for the \lang reduction semantics in Concurrent
Haskell \cite{JonesGF96, MarlowJMR01} to which we can plug-in different
decisions about the non-deterministic transitions of our semantics. Here we
briefly explain the runtime architecture of this interpreter, shown in
\cref{fig:architecture}.

\begin{myfigure*} {\lang runtime architecture.} {fig:architecture} 
\vspace{2ex}
  \begin{center}
    \begin{tikzpicture}
    [node distance=3cm,>=stealth']
      \tikzstyle{supervisor}=[state, draw=blue!50,fill=blue!20, very thick,
      minimum size=3.7em] 
      \tikzstyle{process}=[state, draw=blue!50, fill=blue!20, very thick]
      \tikzstyle{accepting}=[accepting by arrow]
      \tikzstyle{annot}=[rectangle,draw=blue!50,very thick,fill=blue!20]
      \tikzstyle{event}=[rectangle,thick,draw=black!50,fill=yellow!20,text
      width=2cm, text centered,font=\sffamily]

   \tikzstyle{innerWhite} = [semithick, white,line width=2.2pt, shorten >= 4.5pt]
   
      \node[supervisor]                     (s)                       {Sched.};
      \node[supervisor,node distance=7cm]  (g)    [right of=s]        {Gath.}; 
      \node[event]                          (trie) at ($(s)!0.5!(g)$) {Transaction trie};
      
      \node[double arrow, draw, minimum size=1.5cm, minimum width=0.5em]      (a1)
            at ($(s.east)!0.5!(trie.west)$)
            {};

      \node[double arrow, draw, minimum size=1.5cm, minimum width=0.5em]      (a1)
            at ($(g.west)!0.5!(trie.east)$)
            {};

      \node[process]    (p1) [below of=s]       {$ e_1 $}; 
      \node[process]    (p3) [below of=g]       {$ e_n $};
      \node[process]    (p2) at ($(p1)!0.4!(p3)$) {$ e_2 $};
      \node[yshift=-1ex]                  at ($(p2)!0.5!(p3)$) {\textbf{\dots}};
      
      \path[->, dotted, thick]
        (s)  edge [near start, bend left=10] node[xshift=-.4cm]
        {\parbox{2cm}{abort,\\ embed,\\ commit}} 
                                          (p1)
        (s)  edge [bend right=10] node {} (p2)
        (s)  edge [bend right=10] node {} (p3)
        ;
         
      \path[->, dashed, thick]
        (p1)  edge [bend right=10] node {} (g)
        (p2)  edge [bend right=10] node {} (g.235)
        (p3)  edge [near end,bend left=22] 
              node [xshift=1.5cm]
              {\parbox{2.3cm}{side-effect notif.\\\& ack}} (g.250)
        ;
        
    \end{tikzpicture}
  \end{center}
\end{myfigure*} 

The main Haskell threads are shown as round nodes in the figure. Each concurrent
functional expression $e_i$ is interpreted in its own thread according to the
sequential reduction rules in \cref{tab:sequentialReductionRules} of the
previous section. Side-effects in an expression will be generally handled by the
interpreting thread, creating new channels, spawning new threads, and starting
new transactions.

Except for new channel creation, the evaluation of all other side-effects in an
expression will cause a \emph{notification} (shown as dashed arrows in
\cref{tab:sequentialReductionRules}) to be sent to the \emph{gatherer} process
(Gath.). This process is responsible for maintaining a global view of the state of
the running program in a \emph{Trie} data-structure. This data-structure
essentially represents the transactional structure of the program; i.e., the logical
nesting of transactions and processes inside running
transactions:
{\footnotesize
\begin{alltt}
\keyw{data} TTrie = TTrie \{
  threads   :: Set ThreadID,
  children  :: Map TransactionID TTrie, ... \}
\end{alltt}
}

A {\tt TTrie} node represents a transaction, or the top-level of the program.
The main information stored in such a node is the set of threads ({\tt threads})
and transactions ({\tt children}) running in that transactional level. Each
child transaction has its own associated {\tt TTrie} node. An invariant of the
data-structure is that each thread and transaction identifier appears only once
in it. For example the complex program we saw on page \pageref{example:nested}:
$$\nbox{
  \transo{k}{
    (\recvb c; \commitb k)^{\texttt{tid}_1}
    \parallel
    \transo{l}{(\recvb d; \commitb l)^{\texttt{tid}_2}}{()}
    \\
  }{ 
    (); \atomico{l}{\recvb d; \commitb l}{()}
  }
  \qquad\qquad\qquad \parallel
  P^{\texttt{tid}_P}
}$$
will have an associated trie:
{\footnotesize
\begin{alltt}
TTrie\{threads  = \{tid\(_P\)\},
      children = \{\(k\)\,\(\mapsto\)\,TTrie\{threads\, = \{tid\(_1\)\},
                           children = \{\(l\)\,\(\mapsto\)\,TTrie\{threads \,= \{tid\(_2\)\},
                                               \;children = \(\emptyset\)\}\}\}\}\}
\end{alltt}
}

The last ingredient of the runtime implementation is the \emph{scheduler} thread
(Sched. in \cref{fig:architecture}). This makes decisions about the
commit, embed and abort transitions to be performed by the expression threads,
based on the information in the trie. Once such a decision is made by the
scheduler, appropriate signals (implemented using Haskell asynchronous exceptions
\cite{MarlowJMR01}) are sent to the running threads, shown as dotted lines in
\cref{fig:architecture}. Our implementation is parametric to the precise
algorithm that makes scheduler decisions, and in the following section we
describe a number of such algorithms we have tried and evaluated.

A scheduler signal received by a thread will cause the update of the
\emph{local transactional state} of the thread, affecting the future execution
of the thread. The local state of a thread is an object of the {\tt TProcess} data-type:

\begin{center}
\begin{minipage}[t]{.5\linewidth}
\footnotesize
\begin{alltt}
\keyw{data} TProcess = TP \{
  expr :: Expression,
  ctx  :: Context,
  tr   :: [Alternative] \}
\end{alltt}
\end{minipage}
\begin{minipage}[t]{.4\linewidth}
\footnotesize
\begin{alltt}
\keyw{data} Alternative = A \{
  tname :: TransactionID,
  pr    :: TProcess \}
\end{alltt}
\end{minipage}
\end{center}

The local state maintains the expression ({\tt expr}) and evaluation context
({\tt ctx}) currently interpreted by the thread and a list of \emph{alternative}
processes (represented by objects of the {\tt Alternative} data-type). This list
contains the continuations stored when the thread was embedded in transactions. The
nesting of transactions in this list mirrors the transactional nesting in the
global trie and is thus compatible with the transactional nesting of other
expression threads. Let us go back to the example of
page \pageref{example:nested}:
$$\nbox{
  \transo{k}{
    (\recvb c; \commitb k)^{\texttt{tid}_1}
    \parallel
    \transo{l}{(\recvb d; \commitb l)^{\texttt{tid}_2}}{()}
    \\
  }{ 
    (); \atomico{l}{\recvb d; \commitb l}{()}
  }
  \qquad\qquad\qquad \parallel
  P^{\texttt{tid}_P}
}$$
where $P= \sendo{d}{()}; \sendo{c}{()}$. When $P$ is embedded in both $k$ and
$l$, the thread evaluating $P$ will have the local state object
{\footnotesize
\begin{alltt}
TP\{expr = \(P\), tr = \([\)A\{tname = \(l\), \,pr = \(P\)\}, A\{tname = \(k\), pr = \(P\)\}\(]\)\}
\end{alltt}
}
recording the fact that the thread running $P$ is part of the $l$-transaction,
which in turn is inside the $k$-transaction. If either of these transactions
aborts then the thread will rollback to $P$, and the list of alternatives will
be appropriately updated (the aborted transaction will be removed).

Once a transactional reconfiguration is performed by a thread, an
acknowledgment is sent back to the gatherer, who, as we discussed, is
responsible for updating the global transactional structure in the trie. This
closes a cycle of transactional reconfigurations initiated from the process
(by starting a new transaction or thread) or the scheduler (by issuing a commit, embed, or
abort).

What we described so far is a simple architecture for an interpreter of \lang.
Various improvements are possible (e.g., addressing the message bottleneck in the
gatherer) but are beyond the scope of this paper. In the
following section we discuss various policies for the scheduler which we then evaluate
experimentally.


  \section{Transactional Scheduling Policies}         \label{section:strategies}
Our goal here is to investigate schedulers that make decisions on transactional
reconfiguration based only on runtime heuristics. We are currently working on
more advanced schedulers, including schedulers that take advantage of static
information extracted from the program, which we leave for future work.

An important consideration when designing a scheduler is \emph{adequacy}
\cite[Chap. 13, Sec. 4]{Winskel}. 
For a given program, an adequate scheduler is able to produce all outcomes
that the non-deterministic operational semantics can produce for that program.
However, this does \emph{not} mean that the scheduler should be able to produce
all traces of the non-deterministic semantics. Many of these traces will simply
abort and restart the same computations over and over again. Previous work on
the behavioral theory of communicating transactions has shown that all program
outcomes can be reached with traces that \emph{never} restart a computation
\cite{commTransactions}. Thus a goals of our schedulers is to minimize
re-computations by minimizing the number of aborts.

Moreover, as we discussed at the end of \cref{section:language}, many of the
exponential number of embeddings can be avoided without altering the observable
behavior of a program. This can be done by embedding a process inside a
transaction only when this embedding is necessary to enable communication
between the process and the transaction. We take advantage of this in a
\emph{communication-driven} scheduler we describe in this section.

Even after reducing the number of possible non-deterministic choices faced by
the scheduler, in most cases we are still left with a multitude of alternative
transactional reconfiguration options. Some of these are more likely to lead to
efficient traces than other. However, to preserve adequacy we cannot exclude any
of these options since the scheduler has no way to foresee their outcomes. In
these cases we assign different, non-zero probabilities to the available
choices, based on heuristics. This leads to measurable performance improvements,
without violating adequacy. Of course some program outcomes might be more likely
to appear than others. This approach is trading measurable fairness for
performance improvement.

However, the probabilistic approach is \emph{theoretically fair}. Every finite
trace leading to a program outcome has a non-zero probability. Diverging traces
due to sequential reductions also have non-zero probability to occur. The only
traces with zero probability are those in the reduction semantics
that have an infinite number of non-deterministic reductions. Intuitively, these
are unfair traces that abort and restart transactions \emph{ad infinitum}, even if
other options are possible.

\paragraph{Random Scheduler (\defn{R}).}

The very first scheduler we consider is the random scheduler, whose 
policy is to simply, at each point, select one of all the non-deterministic
choices with equal probability, without excluding any of these choices. With
this scheduler any abort, embed, or commit actions are equally likely to happen.
Although this naive scheduler is not particularly efficient, as one would
expect, it is an obviously adequate and fair scheduler according to the
discussion above. If a reduction transition is available infinitely often,
scheduler \defn{R} will eventually select it.

This scheduler leaves much room for improvement. Suppose that a transaction $k$
is ready to commit: 
$$
  \transo{k}{P \parallel \cob k}{Q}
$$
Since \defn{R} makes no distinction between the choices of committing and
aborting $k$, it will often unnecessarily abort $k$. 
All processes embedded in this transaction will have to roll back and
re-execute; if $k$ was a transaction that restarts, the transaction will also
re-execute. This results to a considerable performance penalty. Similarly,
scheduler \defn{R} might preemptively abort a long-running transaction that
could have have committed given enough time and embeddings (for the purpose of
communication).

\paragraph{Staged Scheduler (\defn{S}).}

The staged scheduler partially addresses these issues by prioritizing its
available choices. Whenever a transaction is ready to commit, scheduler
\defn{S} will always decide to send a commit signal to that transaction before
aborting it or embedding another process in it.
This does not violate adequacy; before continuing with the algorithm
of $\defn{S}$, let us examine the adequacy of prioritizing commits over other
transactional actions with an example. 

\begin{examplethm}
Consider the following program in which $k$ is ready to commit.
$$
  \transo{k}{P \parallel \cob k}{Q} \parallel R
$$
If embedding $R$ in $k$ leads to a program outcome, then that outcome
can also be reached after committing $k$ from the residual $P \parallel R$. 

Alternatively, a program outcome could be reachable by aborting $k$ (from the
process $Q \parallel R$). However, the $\cob k$ was spawned from one of the
previous states of the program in the current trace. In that state, transaction
$k$ necessarily had the form:
$
  \transo{k}{P' \parallel E[\commitb k]}{Q}
$.
In that state the abort of $k$ was enabled. Therefore, the staged interpreter
indeed allows a trace leading to the program state $Q \parallel R$ from which
the outcome in question is reachable.
\qed
\end{examplethm}

If no commit is possible for a transaction, the staged interpreter prioritizes
embeds into that transaction over aborting the transaction. This is again an
adequate decision because the transactions that can take an abort reduction
before an embed step have an equivalent abort reduction after that step.

When no commit nor embed options are available for a transaction, the staged
interpreter lets the transaction run with probability $0.95$, giving more
chances to make progress in the current trace, and with probability $0.05$ it
aborts it---these numbers have been fine-tuned with a number of experiments.

The benefit of the heuristic implemented in this scheduler is that it minimizes
unnecessary aborts improving performance. Its drawback is that it does not abort
transactions often, thus program outcomes reachable only from transactional
alternatives are less likely to appear. Moreover, this scheduler does not avoid
\emph{unnecessary embeddings}.

\paragraph{Communication-Driven Scheduler (\defn{CD}).}

To avoid spurious embeddings, scheduler \defn{CD} improves over \defn{R} by
performing an embed transition only if it is \emph{necessary} for an imminent
communication. For example, in the following program state the embedding
of the right-hand-side process into $k$ will never be chosen.
$$
  \transo{k}{E[\recvb c]}{Q} \parallel ((); \sendo{c}{v})
$$
However, after that process reduces to an output, its embedding into $k$ will be
enabled. Because of the equivalence
$$
  \transo{k}{P}{Q} \parallel R \equiv_{\text{cxt}} \transo{k}{P \parallel R}{Q \parallel R}
$$
which we previously discussed, this scheduler is adequate.

For the implementation of this scheduler we augment the information stored in the
trie data-structure (\cref{section:implementation}) with the channel which each
thread is waiting to communicate on (if any).

As we will see in \cref{section:experiments}, this heuristic significantly
boosts performance because it greatly reduces the exponential
number of embedding choices.

\paragraph{Delayed-Aborts Scheduler (\defn{DA}).}

The final scheduler we report is \defn{DA}, which adds a minor
improvement upon scheduler \defn{CD}. This scheduler keeps a timer for each
running transaction $k$ in the transaction trie. This timer is reset whenever a
communication or transactional operation happens inside $k$. Transaction $k$
will only be considered for an abort when this timer expires. This strategy
benefits long-running transactions that perform multiple communications before
committing. The \defn{CD} scheduler is obviously adequate because it only adds
time delays.


\section{Evaluation of the Interpreters}                   \label{section:experiments}

We now report the experimental evaluation of interpreters using the preceding
Scheduling policies. The interpreters were compiled with GHC 7.0.3, and the
experiments were performed on a Windows 7 machine with 
Intel$\textsuperscript{\textregistered}$ Core\texttrademark i5-2520M 250
GHz processor and 8Gb of RAM.
We run several versions of two programs:
\begin{enumerate}
  \item The three-way rendezvous (3WR) in which a number of processes compete to
    synchronize on a channel with \emph{two} other processes, forming groups of
    three which then exchange values. This is a standard
    example of multi-party agreement \cite{reppy, tevents, composableMemoryTransactions}. 
    In the \lang implementation of this example each process
    nondeterministically chooses between being a \emph{leader} or
    \emph{follower} within a communicating transaction. If a leader and two
    followers communicate, they can all exchange values and commit; any other
    situation leads to deadlock and eventually to an abort of some of the
    transactions involved.

  \item The SNO example of the introduction, as implemented in
    \cref{section:language}, with multiple instances of the \Alice, \Bob,
    \Carol, and \David processes.
\end{enumerate}
To test the scalability of our schedulers, we tested a number of versions of
the above programs, each with a different number of competing parallel
processes. Each process in these programs continuously performs 3WR 
or SNO cycles and our interpreters are instrumented to measure the number of
operations in a given time, from which we compute the \emph{mean throughput} of
successful 3WR or SNO operations. The results are shown in
\cref{fig:experiments}.

\begin{myfigure}{Experimental Results.}{fig:experiments} 
\pgfplotsset{
    every axis/.append style={
        scale only axis,
        width=0.38\textwidth,
        xtick={5,10,15,20},
    }
}
  \begin{tabular}{p{.50\linewidth}@{\quad}p{.48\linewidth}}
\tikz{  
  \begin{semilogyaxis}[ title={Three-Way Rendezvous},ylabel={ Committed ops/second},
      legend entries={\defn{R},\defn{S},\defn{CD},\defn{TA},\defn{ID}},
      legend columns=-1,
      legend to name=named
    ] 
  \addplot coordinates{ 
    (3, 0.24)  
    (6, 0.003) 
    (9, 0.015) 
    (12, 0) 
    (15, 0)
    (18, 0)
  };
  \addplot coordinates{
    (3,  12.23666667)
    (6,  0.066666667)
    (9,  0.044444444)
    (12, 0.166666667)
    (15, 0)
    (18, 0)
  };
  \addplot coordinates{ 
    (3,  21.154)
    (6,  10.97)
    (9,  2.183)
    (12, 0.837)
    (15, 0.22)
    (18, 0.127)
  };
  \addplot coordinates{ 
    (3,  30.78)
    (6,  9.487)
    (9,  1.017)
    (12, 0.22)
    (15, 0.09)
    (18, 0.014)    
  };
  \addplot coordinates{
    (3,  100.9952381)
    (6,  141.3952381)
    (9,  175.2166667)
    (12, 207.5222222)
    (15, 236.9555556)
    (18, 219.2666667)
  };
  \end{semilogyaxis} 
} 
&
\tikz{ 
  \begin{semilogyaxis}[ title={SNO Example} ] 
    \addplot coordinates{ 
      (4,  0.25)
      (8,  0.38)
      (12, 0.326666667)
      (16, 0.136666667)
      (20, 0.19047619)
      (24, 0.06)
    };
    \addplot coordinates{
      (4,  1.76)
      (8,  8.936666667)
      (12, 2.49)
      (16, 2.07)
      (20, 0.508333333)
      (24, 0.406666667)
    };
  \addplot coordinates{
    (4,  3.196666667)
    (8,  9.813333333)
    (12, 1.753333333)
    (16, 1.493333333)
    (20, 0.55)
    (24, 0.39)
  };
  \addplot coordinates{
    (4,  4.45)
    (8,  10.69666667)
    (12, 0.73)
    (16, 0.953333333)
    (20, 0.104166667)
    (24, 0.116666667)
  };
  \addplot coordinates{
    (4,  100.9952381)
    (8,  161.2)
    (12, 121.4)
    (16, 246.94)
    (20, 256.2)
    (24, 262.9333333)
  };
  \end{semilogyaxis} 
} 
\\

\qquad\qquad~  Number of concurrent processes 
\hfill
&
\hfill\raisebox{-1em}{\ref{named}}~~~
\\[1em]
  \end{tabular}
\end{myfigure} 

Each graph in the figure contains the mean throughput of operations (in
logarithmic scale) as a function of the number of competing concurrent \lang
processes. The graphs contain runs with each scheduler we discussed (random
\defn{R}, staged \defn{S}, communication-driven, \defn{CD}, and
communication-driven with timed aborts \defn{TA}) as well as with an
\emph{ideal} non-transactional program (\defn{ID}). The ideal program in the
case of 3WR is similar to the TCML, non-abstract implementation \cite{reppy}.
The ideal version of the SNO is running a simpler instance of the scenario,
without any \Carol processes---this instance has no deadlocks and therefore
needs no error handling. Ideal programs give us a performance upper
bound.

As predictable, the random scheduler (\defn{R})'s performance is the
worst; in many cases \defn{R} could not perform any operations in the
window of measurements (30sec).

The other schedulers perform better than \defn{R} by an order of
magnitude. Even just prioritizing the transactional reconfiguration
choices significantly cuts down the exponential number of inefficient traces.
However, none of the schedulers scale to programs with more processes; their
performance deteriorates exponentially. In fact, when we go from the
communication-driven (\defn{CD}) to the timed-aborts (\defn{TA}) scheduler we
see worst throughput in larger process pools. This is because with
many competing processes there is more possibility to enter a path to
deadlock; in these cases the results suggest that it is better to abort early.

The upper bound in the performance, as shown by the throughput of \defn{ID} is
one order of magnitude above that of the best interpreter, when there are few
concurrent processes, and (within the range of our experiments) two orders when
there are many concurrent processes. The performance of \defn{ID} is increasing
with more processes due to better utilization of the processor cores.

It is clear that in order to achieve a pragmatic implementation of \lang we need
to address the exponential nature in consensus scenarios as the ones we tested
here. Our exploration of purely runtime heuristics shows that performance can be
improved, but we need to turn to a different approach to close the gap between
ideal ad-hoc implementations and abstract \lang implementations.


\section{Conclusions and Future Work} \label{section:related-work}  

Consensus is an often occurring problem in concurrent and distributed
programming.
The need for developing programming language support for consensus has already
been identified in previous work on \emph{transactional events} (TE)
\cite{tevents}, \emph{communicating memory transactions} (CMT) \cite{cmt},
\emph{transactors} \cite{transactors} and \emph{cJoin} \cite{cjoin}. These
approaches propose forms of restarting communicating transactions, similar to
those described in \cref{section:language}. TE, CMT and Transactors can be used
to implement the instance of the Saturday Night Out (SNO) example in this paper.
TE extends CML events with a transactional sequencing operator; transactional
communication is resolved at runtime by search threads which exhaustively
explore all possibilities of synchronization, avoiding runtime aborts.
CMT extends STM with asynchronous communication, maintaining a directed
dependency graph mirroring communication between transactions; STM abort
triggers cascading aborts to transactions that have received values from
aborting transactions.
Transactors extend actor semantics with fault-tolerance primitives, enabling the
composition of systems with consistent distributed state via distributed
checkpointing.
The cJoin calculus extends  the Join calculus with isolated transactions which
can be merged. Merging and aborting are managed by the programmer, offering a
manual alternative to TCML's nondeterministic transactional operations.  It is
unclear to us how to write a straightforward implementation of the SNO example
in cJoin.
Reference implementations have been developed for TE, CMT and cJoin. The
discovery of efficient implementations for communicating transactions
could be equally beneficial for all approaches.
Stabilizers \cite{stabilizers} add transactional support for fault-tolerance in the
presence of transient faults but do not directly address concensus scenarios such as the
SNO example.

This paper presented \lang, a simple functional language with
build-in support for consensus via communicating transactions. This is a
construct with a robust behavioral theory supporting its use as a programming
language abstraction for automatic error recovery
\cite{commTransactions,livenessComm}. TCML has a simple
operational semantics and can simplify the programming of advanced consensus scenarios; 
we introduced such an example (SNO) which has a natural
encoding in \lang.

The usefulness of communicating transactions in real-world applications,
however, depends on the invention of efficient implementations. This paper
described the obstacles we need to overcome and our first results in a recently
started project on developing such implementations. We gave a framework to
develop and evaluate current and future runtime schedulers of communicating
transactions and used it to examine schedulers which are based solely on runtime
heuristics. We have found that some heuristics improve upon the performance of a
naive randomized implementation but do not scale to programs with significant
contention, where an exponential number of alternative computation paths lead to
necessary rollbacks. It is clear that purely dynamic strategies do not lead to
sustainable performance improvements.

In future work we intend to pursue a direction based on the extraction of
information from the source code which will guide the language runtime. This
information will include an abstract model of the communication behavior of
processes that can be used to predict with high probability their future
communication pattern. A promising approach to achieve this is the development
of technology in type and effect systems and static analysis. Although the
scheduling of communicating transactions is theoretically computationally
expensive, realistic performance in many programming scenarios could be
achievable.

%
%

\bibliographystyle{splncs03}
\bibliography{paper}

\begin{thebibliography}{10}
\providecommand{\url}[1]{\texttt{#1}}
\providecommand{\urlprefix}{URL }

\bibitem{cjoin}
Bruni, R., Melgratti, H., Montanari, U.: cjoin: Join with communicating
  transactions. To appear in MSCS

\bibitem{livenessComm}
De~Vries, E., Koutavas, V., Hennessy, M.: Liveness of communicating
  transactions. pp. 392--407. APLAS'10

\bibitem{tevents}
Donnelly, K., Fluet, M.: Transactional events. pp. 124--135. ICFP '06

\bibitem{transactors}
Field, J., Varela, C.A.: Transactors: A programming model for maintaining
  globally consistent distributed state in unreliable environments. POPL '05

\bibitem{composableMemoryTransactions}
Harris, T., Marlow, S., Jones, S.P.L., Herlihy, M.: Composable memory
  transactions. Commun. ACM pp. 91--100 (2008)

\bibitem{herlihybook}
Herlihy, M., Shavit, N.: The art of multiprocessor programming. Kaufmann (2008)

\bibitem{JonesGF96}
Jones, S.P.L., Gordon, A.D., Finne, S.: Concurrent haskell. POPL'96

\bibitem{distribComp}
Kshemkalyani, A.D., Singhal, M.: Distributed Computing: Principles, Algorithms,
  and Systems. Cambridge University Press (2008)

\bibitem{cmt}
Lesani, M., Palsberg, J.: Communicating memory transactions. PPoPP '11

\bibitem{MarlowJMR01}
Marlow, S., Jones, S.P.L., Moran, A., Reppy, J.H.: Asynchronous exceptions in
  haskell. PLDI'01

\bibitem{reppy}
Reppy, J.H.: Concurrent programming in ML. Cambridge University Press (1999)

\bibitem{tcml-report}
Spaccasassi, C.: Transactional concurrent ml. Tech. Rep. TCD-CS-2013-01

\bibitem{commTransactions}
de~Vries, E., Koutavas, V., Hennessy, M.: Communicating transactions. pp.
  569--583. CONCUR'10

\bibitem{stabilizers}
Ziarek, L., Schatz, P., Jagannathan, S.: Stabilizers: a modular checkpointing
  abstraction for concurrent functional programs. ICFP '06

\end{thebibliography}

\end{document}